\documentstyle[12pt]{article}

\begin{document}

\title{An exact derivation of the dissipation rate correlation 
exponent $\mu$ in fully-developed turbulence}

\author{  C. Jayaprakash and F. Hayot\\ Department of Physics, \\
The Ohio State University, \\ Columbus, Ohio 43210}

\maketitle
\begin{abstract}

We derive for the Navier-Stokes equation an
exact equation satisfied by the dissipation rate correlation
function, $\langle\epsilon(\vec{x}+\vec{r},t+\tau)\epsilon(
\vec{x},t)\rangle$.  In the equal time limit, for the homogeneous,
isotropic state of fully-developed turbulence, we show that the correlation
function behaves as $Ar^{-\mu_1}\,+\,Br^{-\mu_2}$ with 
$\mu_1=2-\zeta_6$ and
$\mu_2=z_4''-\zeta_4$ for $r$ in the inertial range;  the $\zeta$'s are exponents of
velocity structure functions and $z_4''$ is a dynamical
exponent characterizing the $4$th order dynamical structure function. This
provides the first direct derivation of the exponents of the dissipation-rate
correlation.
\end{abstract}

\noindent{PACS numbers: 0.5.45.+b, 47.10.+g}

\vspace{0.5cm} \pagebreak The statistical properties of the energy
dissipation rate $\epsilon(\vec x,t)$ defined by \begin{equation}
\epsilon(\vec x,t)\,=\,\frac{\nu}{2}\,\sum_{i,j}\,
(\,\partial_iu_j\,+\,\partial_ju_i\,)^2 \end{equation} where
$\partial_i\,=\,\partial/\partial x_i$ have played a crucial role
in our understanding of fully-developed 
turbulence in  incompressible fluids.\cite{frisch,my}  In the original Kolmogorov theory
$\epsilon$ is replaced by $\langle \epsilon\rangle$ and the
spatial fluctuations are ignored; the effect of fluctuations of
$\epsilon$ pointed out by Landau have been explored in the context
of the lognormal model and its multifractal
generalizations.\cite{sreeni1}  The intermittent behavior of
turbulent fluctuations is reflected in the power-law behavior of
the correlations of $\epsilon$:
\begin{equation} \langle\,\epsilon(\vec x)\,\epsilon(\vec x +\vec
r)\,\rangle\,\sim\,(r/L)^{-\mu} \end{equation} where $L$ is a
length  scale characteristic of the large-scale flow and 
$r=\vert \vec r\vert$ belongs to the inertial range.  Simple
dimensional analysis, noting that the dimension of $\epsilon$ is
$V^3/L$, yields the identification $\mu\,=\,2\,-\,\zeta_6$; the
exponents of the $q^{th}$-order (longitudinal) structure function,
$\zeta_q$ are defined by
\begin{equation}
\label{zetaq}
S_q\,\equiv\,\langle\,[\,\delta \vec u \cdot
\hat{r}\,]^q\,\rangle\,\sim\,(r/L)^{\zeta_q}\end{equation} 
where
$\delta \vec u \,=\,\vec u(\vec x+\vec r,t)-\vec u(\vec x,t)$.
Within the original Kolmogorov theory $\zeta_6\,=\,2$ and
consequently $\mu\,=\,0$. Thus the deviation of $\mu$ from zero
is a measure of the degree of intermittency and is an important
quantity for understanding fully developed turbulence. This
breakdown of simple Kolmogorov scaling has been studied
experimentally and a review of the experiments\cite{sreeni2} gives
a ``best" estimate for $\mu$ of $0.25 \pm 0.05$  which is
consistent with the experimentally measured value of $\zeta_6$. In
this Letter we provide a simple and direct derivation of the
values of $\mu$ for the Navier-Stokes equation
by deriving the exact equation satisfied by the dissipation rate
correlations. The equation for the dissipation rate contains two
contributions, one a second spatial derivative of an appropriate
sixth-order structure function and another a second temporal
derivative of a fourth-order structure function; there are, in
addition, pressure-dependent terms and no other velocity-dependent
terms. The use of {\em dynamical} structure functions, i.e., $S_q$ in Eqn.
(\ref{zetaq}) defined with 
 $\delta \vec u \,=\,\vec u(\vec x+\vec r,t+\tau)-\vec
u(\vec x,t)$, is key to our derivation. Both spatial and temporal
derivatives of the dynamic structure functions occur naturally and
the equal time limit of the derivatives is related to the
correlation of the energy dissipation rates. We illustrate the
method first with an application to the $1d$ stochastic Burgers
equation\cite{burgers} which provides an extremely fruitful model
for elucidating some of the conceptual  and mathematical features
of $3d$ turbulence. After providing the derivation in the
Navier-Stokes case we comment on  our results  and discuss
some experimental implications. 

{\em Derivation  for the stochastic Burgers equation: } We discuss
first the Burgers equation (describing a one-dimensional, 
compressible fluid without pressure) driven by a stochastic force
$f(x,t)$:
$$ \partial u/\partial t\,+\,u \partial u  /\partial x
~=~ \nu \partial^2u/\partial x \,+\,  f ~~.$$  The stochastic
force is a Gaussian white-noise with algebraic spatial
correlations given in $k$-space by \begin{equation} \label{noise}
\langle\hat{f}(k,t)\hat{f}(k',t')\rangle\,
=\,D_0|k|^{\beta}\,\delta_{k+k',0}\delta(t-t')
\end{equation} with $-1 \le \beta< 0$. For these values of $\beta$, 
intermittent behavior of  the structure functions has been shown
to occur.\cite{hj1}  We will first illustrate our approach by doing
the simplest calculation. We employ the notation $u=u(x,t)$,
$u'=u(x',t')$ where $x=R+r/2$, $x'=R-r/2$, $t=T+\tau/2$, and
$t'=T-\tau/2$. We wish to find an equation for $\langle \epsilon
\epsilon'\rangle$ where $\epsilon\,=\,\nu(\partial u/\partial
x)^2$ is the dissipation rate. Multiplying the Burgers equation by
$u$ we have 
$$\nu u
\partial^2 u/\partial x\,+\,fu\,=\,u\partial
u/\partial t\,+\,u^2\partial u/\partial x~$$
and the corresponding equation for $u'$. The idea
is simply to multiply the two equations to obtain
\begin{equation}
\label{basicburg0}
[\nu u\partial^2 u/\partial x\,+\,fu]\,
[\nu u'\partial^2 u'/\partial x'\,+\,f'u']\,=\,
[u\partial u/\partial t\,+\,u^2\partial u/\partial x]\,
[u'\partial u'/\partial t'\,+\,u'^2\partial u'/\partial x']
\end{equation}
and average over the
homogeneous, steady state of Burgers turbulence; the identity $\nu
u\partial^2 u/\partial x^2\,=\,\nu\partial^2 (u^2/2)/\partial
x^2\,-\,\epsilon$ is crucial. We obtain after some straightforward
rearrangements of the terms on the right-hand side using
$\partial/\partial t\,=\,\partial/\partial_\tau$ and
$\partial/\partial t'\,=\,-\partial/\partial_\tau$ when acting on
averages in the steady state,
\begin{equation} 
\label{burgfin0}
\langle \epsilon
\epsilon'\,\rangle\,-\,\langle \epsilon \rangle^2\,\approx\,
-\frac{1}{4}\frac{\partial^2 }{\partial \tau^2}\langle
u^2u'^2\rangle \,-\,\frac{1}{6}\frac{\partial^2 }{\partial \tau
\partial r}\langle (u+u')u^2u'^2\rangle\,-\,\frac{1}{9}
\frac{\partial^2}{\partial r^2}\langle u^3u'^3\rangle
\end{equation}
 All the terms in the above equation
are  assumed to be evaluated in the limit $\tau \rightarrow 0^+$. 
The {\em only} reason for the approximate sign is that we have not
displayed terms which depend explicitly on $\nu$ or the noise  and are
negligible in the inertial range as $\nu \rightarrow 0$: for 
example,
a term such as $\nu^2\,\partial_r^4 S_4$ is negligible 
since $S_4$ is finite. 
In  the equal-time limit the noise-velocity correlations that occur 
on the left-hand side of Eqn. (\ref{basicburg0}) 
can be evaluated 
by using the Donsker-Novikov-Varadhan
result\cite{frisch43}; of these the only term that does not
vanish in the inertial range is the subtracted term $\langle
\epsilon\rangle^2$ in Eqn. (\ref{burgfin0}).

We immediately observe  that we obtain the second temporal derivative
of a fourth-order structure function and the second spatial
derivative of a sixth-order structure function which yield the two exponents
referred to in the abstract.
 Of course, the equation is not
manifestly form-invariant under Galilean transformations and we
have to understand the role of the cross spatio-temporal
derivative. We address this issue using a different
version of the above equation. 

A more elegant form can be obtained by multiplying Burgers
equations  for $u$ and $u'$ by  $\delta u = u-u'\equiv
u(x,t)-u(x',t')$, multiplying the two equations  and averaging over the
homogeneous turbulent state as before. We obtain
\begin{equation}\label{basicburg}
\langle\,\delta u (\,\nu
\partial^2 u\,+\,f\,)\delta u (\,\nu
\partial'^2u'\,+\,f')\,\rangle\,\,=\,
\langle\, \delta u\,\frac{Du}{Dt}\,\delta
u\,\frac{Du'}{Dt'}\,\rangle~\end{equation} where $D/Dt$ represents
the convective derivative, i.e., $Du/Dt\,=\,\partial u/\partial t
\,+\,u\partial u/\partial x$,  and we have used the shorthand notation 
$\partial \equiv
\partial/\partial x$ and $\partial'\equiv \partial/\partial x'$. Evaluating the
purely viscous terms on the
left-hand side of Eq. (\ref{basicburg}) yields
$$\nu^2\,\left\langle (\delta u)^2 \partial^2u\,\partial'^2 u'
\,\right \rangle\,=\,-2\langle\,\epsilon\epsilon'\,\rangle
\,+\,\frac{\nu^2}{12}\partial_r^4\,\langle\, (\delta u)^4\,\rangle
\,+\,\nu\,\partial_r^2\langle\,(\epsilon+\epsilon') (\delta
u)^2\,\rangle $$ where the only non-vanishing term in the 
inertial
range  is $-2\langle\,\epsilon\epsilon'\,\rangle$ in the limit 
$\nu \rightarrow 0$ since as before the rest are
products of powers of $\nu$ and finite correlation functions.  The noise
terms can again be evaluated in the $\tau \rightarrow 0^+$ limit
to yield $2\langle \epsilon \rangle^2\,-\,(1/72)(\partial
S_3/\partial r)^2$ apart from terms which vanish as $\nu
\rightarrow 0$. The right-hand side of Eq. (\ref{basicburg})
contains four terms and their  evaluation is facilitated by
kinematic results which can be obtained in a straightforward way
such as
\begin{eqnarray*}
\partial_\tau^2\,\langle \,(\delta u)^4\,\rangle&=&
12\,\langle\,(\delta u)^2\,\partial_{t}u\,\partial_{t'}u'
\,\rangle ~~~~~~~~~~~~(8a) \\
\partial_r^2\,\langle(\delta u)^6\,\rangle&=& 30\,\langle\,(\delta
u)^4\,\partial_xu\partial_{x'}u'\,\rangle ~~~~~~~~~~(8b)\\
\partial_r^2\,\langle(u+u')^2 (\delta u)^4\rangle&=&
\langle[-2(\delta u)^4+12(u+u')^2(\delta u)^2]
\partial_xu\partial_{x'}u'\rangle ~~~~~~~~~~(8c)
\end{eqnarray*}
\addtocounter{equation}{1} These enable one to simplify the temporal 
and spatial second derivatives. The term involving 
cross (space and time) derivatives  can be simplified by employing
\begin{eqnarray*}
\partial_\tau\partial_r\,\langle \,(\delta u)^4u\,\rangle&=&
\langle\,[12 (\delta u)^2u\,+\,4(\delta u)^3]\partial_{x}u\,\partial_{t'}u'
\,\rangle ~~~~~~~~~~~~(9a) \\
\partial_\tau\partial_r\,\langle \,(\delta u)^4u'\,\rangle&=&
\langle\,[12 (\delta u)^2u'\,-\,4(\delta u)^3]\partial_{x'}u'\,\partial_{t}u
\,\rangle ~~~~~~~~~~~~(9b) \\
\partial_\tau\partial_r\,\langle \,(\delta u)^5\,\rangle&=&
20\langle(\delta u)^3]\partial_{x}u\,\partial_{t'}u'
\rangle\,=\,
20\langle(\delta u)^3]\partial_{x'}u'\,\partial_{t}u'\rangle\,.
~~~~~~~~~~~(9c)
\end{eqnarray*}
\addtocounter{equation}{1}
We obtain after a few manipulations
\begin{equation}
\label{burgfund} \langle\,\epsilon\epsilon'\,\rangle\,-\,\langle
\epsilon \rangle^2
\,\approx\,-\frac{1}{24}\,[\,\partial_\tau^2\,S_4
\,+\,\partial_\tau\partial_r\,S_{4,1}\,+\,\frac{1}{4}\partial_r^2
S_{4,2} \,]
\,+\,\frac{1}{288}\partial_r^2\,S_6\,-\,\frac{1}{144}(\partial_r
S_3)^2
\end{equation}
where we have used the notation
$S_{p,q}\,=\,\langle\,(u-u')^p\,(u+u')^q\,\rangle$. Note that
$\partial_\tau^2\,S_4
\,+\,\partial_\tau\partial_r\,S_{4,1}\,+\,\frac{1}{4}\partial_r^2
S_{4,2}$ is  the convective second derivative  of the fourth-order
structure function, $D^2S_4/D\tau^2$. This term is manifestly
Galilean-invariant as is evident if we recall that $\partial_\tau
\langle\, (u-u')^n \rangle\,+(1/2)\partial_r \langle \,(u+u')
(u-u')^n\, \rangle$ is Galilean-invariant. The last term on the
right-hand side which arises from the noise-velocity 
correlations leads to
$r^{2\zeta_3-2}$ which yields for the dissipation-rate 
correlation exponent precisely the scaling value of 
$2\,-\,\zeta_6$; the explicit value is  given by 
$2-2\zeta_3\,=\,2+2\beta$ 
where the last equality follows from 
the von-Karman-Howarth relation derived previously by
us\cite{hj3}, $\zeta_3\,=\,-\beta$.
In the multifractal regime this term is, of 
course, subdominant. The crucial
feature of Eq. (\ref{burgfund})  is that the non-scaling behavior
of the dissipation rate correlations is determined by (1)
$\partial_r^2S_6$ term which leads to $\mu_1\,=\,2\,-\,\zeta_6$ and
(2)$D ^2S_4/D\tau^2$. Let us define $z''_4$ to be the dynamical
scaling exponent describing the $\tau \rightarrow 0$ limit of the
Galilean-invariant, convective second derivative of $S_4$. This yields another
possible inertial range behavior for $\langle \epsilon \epsilon'\rangle$
with  $\mu_2\,=\,z''_4\,-\,\zeta_4$.   The 
need for a sequence of  dynamical exponents is a consequence  of 
the occurrence
of temporal multiscaling in the dynamical structure functions as has been
emphasized earlier\cite{lpp}; thus  different order temporal
derivatives of $S_p(r,\tau)$ can lead to different dynamical
exponents. In dealing with dynamic structure functions it is
important to recall that we have used the Eulerian description;
therefore,  ordinary dynamic scaling and a fortiori, dynamic
multifractality, are complicated by the presence of sweeping
terms. For example, in $S_p(r=0,\tau)$ (obtained from 
measurements of velocity differences at a given point 
at finite values of the time difference) 
the kinematic exponent $z=1$ arising from sweeping
occurs.\cite{tennekes} However, in the (Galilean invariant)
convective derivative which occurs in the equation above in the
$\tau \rightarrow 0$ limit only the intrinsic, dynamical exponent
occurs. Thus the exact equation neatly
picks out the intrinsic dynamical exponent. We will consider the implications
of the two intermittency exponents after the corresponding derivation for the
three-dimensional problem.

{\bf Derivation for the Navier-Stokes equation:} We 
consider the Navier-Stokes
equation for the velocity field $\vec u(\vec x ,t)$ driven by a
stochastic driving force $\vec f(\vec x, t)$ with zero mean and
variance given by
\begin{equation}
\langle \, \hat{f}_i(\vec k) \hat{f}_j(\vec k') \, \,\rangle ~=~
P_{ij}(\vec k)~ D(\vec{k})
\delta_{\vec{k}+\vec{k}',\vec{0}}\,\delta(t-t')
\end{equation}
where $P_{ij}(\vec k)$ is the transverse projection operator given
by $\delta_{ij}-(k_ik_j/k^2)$. The noise covariance $D(\vec k)$ is
assumed to be peaked around $k_0\,\sim 1/L$ with a  narrow width.
In contrast to the Burgers equation the detailed form of the noise 
correlation is not crucial in the $3d$ problem; the noise maintains a
fully-developed turbulent state and allows one to define averages as 
noise ensemble averages. 
We will find it useful to define the quantity
\begin{equation}
\epsilon_{ij}\,\equiv\,\nu\,\partial_\ell
u_i\,\partial_\ell\,u_j~.\end{equation} The dissipation rate
$\epsilon$ (cf. Eq. (1)) of
 an incompressible fluid obeys the relation \begin{equation}
\epsilon\:=\:\epsilon_{ii}\,-\,\nu \nabla^2
\tilde{p}~~\end{equation} where $\tilde p\,=\,p/\rho$ ($\rho$ is
the constant density) and the summation convention of
 summing over repeated indices is used. We remark that
$\langle \epsilon_{ij}\rangle\,\propto\,\delta_{ij}$ in isotropic
turbulence. 

We use the notation $\vec x\,=\,\vec R\,+\,(1/2)\vec r,~~
t\,=\,T\,+\,(1/2)\tau $ and $\vec x'\,=\,\vec R\,-\,(1/2)\vec r,
~~t'\,=\,T\,-\,(1/2)\tau$. Multiplying  the Navier-Stokes equation
for $u_i\,=\,u_i(\vec x, t)$ by $u_i$ and summing  over $i$ and
using $\nu u_i \nabla^2 u_i\,=\,\nu
\nabla^2(u^2/2)\,-\,\epsilon_{ii}$ one finds
\begin{equation} \label{nseq} -\epsilon_{ii}\,+\,\nu\nabla^2 (u^2/2)
\,+\,\vec{f}\cdot \vec{u}\,=\, u_i\partial_t
u_i\,+\,u_iu_l\partial_l u_i+u_i\partial_i\tilde{p}\,
~~.\end{equation}
 Again we
write a similar equation for $u'_i\,=\,u_i(\vec x',t')$ and
multiply the two equations and average over the homogenous, steady
state of isotropic turbulence leading to
\begin{eqnarray}
&&\langle\,\epsilon_{ii}\epsilon'_{jj}\,\rangle\,-\,
\frac{\nu}{2}\nabla^2_r\,\langle\,\epsilon_{ii}u'^2+\epsilon'_{ii}u^2\,\rangle
\,+\,\frac{\nu^2}{4}\nabla^2_r\nabla^2_r\langle\,u^2u'^2\,\rangle
\,+\,\mbox{noise terms} \nonumber \\
&=& -\frac{1}{4}\frac{\partial^2}{\partial
\tau^2}\langle\,u^2u'^2\,\rangle\,-\,\frac{1}{2}\frac{\partial^2}{\partial
\tau \partial
r_i}\langle\,(u_i+u'_i)u^2u'^2\,\rangle\,-\,\frac{1}{4}\frac{\partial^2}
{\partial r_i \partial r_j}\,\langle\,u_iu_ju^2u'^2\,\rangle
\nonumber \\
&& \,+\,\mbox{pressure terms}~~.
\end{eqnarray}
The second and third terms on the left-hand side are negligible in
the inertial range and will be suppressed hereafter; the velocity
terms clearly have the same form as in the Burgers equation
consisting of a second derivative with respect to $r$ of the
sixth-order structure function and a second derivative with
respect to $\tau$ of a fourth-order structure function. We have
not displayed the noise and pressure terms since this calculation
merely illustrates the simple steps involved in the derivation. We
will discuss
them in the Galilean-invariant form given next.

The manifestly Galilean-invariant form is somewhat more difficult
to obtain in the $3-d$ problem. We proceed from the Navier-Stokes
equation in two different ways and add the results appropriately.
First multiply the Navier-Stokes equation for $u_i$ and $u'_i$ by $\delta
u_i\,\equiv\,u_i(\vec x,t)\,-\,\vec u_i(\vec x',t')$ and multiply the two
equations to obtain 
\begin{equation}
\label{nsfirst}
\delta u_i\,(\nu \nabla^2 u_i\,+\,f_i)\,
\delta u_j\,(\nu \nabla'^2 u'_j\,+\,f'_j\,)
\:=\:\delta u_i\,(\,
\frac{Du_i}{Dt}\,+\,\partial_i\tilde{p}\,)\,
\delta u_j\,(\,
\frac{Du'_j}{Dt'}\,+\,\partial'_j\tilde{p'}\,)
\:\:.\end{equation} 
We now average this equation over the isotropic, 
homogeneous, steady state of turbulence.  On the left-hand side we
obtain, in addition to the noise terms,
$-\,\langle\,\epsilon_{ii}\epsilon'_{jj}\,+\,
\epsilon_{ij}\epsilon'_{ij}\,\rangle$ as the only terms which
survive in the inertial range. We do a similar set of
manipulations with the Navier-Stokes equation for $u_i$ multiplied
by $\delta u_j$ and obtain
\begin{equation}
\label{nssecond}
\delta u_j\,(\nu \nabla^2 u_i\,+\,f_i)\,
\delta u_j\,(\nu \nabla'^2 u'_i\,+\,f'_i\,)
\:=\:\delta u_j\,(\,
\frac{Du_i}{Dt}\,+\,\partial_i\tilde{p}\,)\,
\delta u_j\,(\,
\frac{Du'_i}{Dt'}\,+\,\partial'_i\tilde{p'}\,)
\:\:.\end{equation} 
We 
add Eqn. (\ref{nsfirst}) to twice Eqn. (\ref{nssecond}) 
and average the resulting equation 
over the turbulent  state. The
right-hand side of the resultant equation includes, apart from the
pressure terms,
$$\langle\,\delta u_i \frac{Du_i}{Dt}\delta u_j
\frac{Du'_j}{Dt'}\rangle~+~
 2\langle\,\delta u_j \frac{Du_i}{Dt}\delta u_j
\frac{Du'_i}{Dt'}\rangle\:\:.$$ We can show after some algebraic
manipulations involving kinematic relations which are the
three-dimensional generalizations of those displayed in Equations
(8) and (9)  that these terms correspond to $$
(1/16)(\partial^2/ \partial r_i
\partial r_j)\,\langle\,\delta u_i \delta u_j (\delta \vec u \cdot
\delta \vec u)^2 \,\rangle\,-\,(1/4)(D^2/D\tau^2)
\langle\,(\delta \vec u \cdot \delta \vec u)^2 \,\rangle\,     $$
where $D^2/D\tau^2\langle f\rangle\,=\,\partial^2/\partial \tau^2 
\langle f\rangle\,+\,\partial^2/\partial \tau \partial r_i\langle
(u_i+u_i')f\rangle\,+\,(1/4)\partial^2/\partial r_i\partial r_j 
\langle (u_i+u'_i)(u_j+u'_j)f\rangle$. 
This shows the occurrence of the two key structure function derivatives.
We have evaluated the pressure terms  and  find
\begin{eqnarray}
\label{nsbasic}
&&2\langle\,\epsilon_{ii}\epsilon'_{jj}\,+\,2\epsilon_{ij}\epsilon'_{ij}\,\rangle
\,+\,\mbox{noise terms}\,=\, \frac{1}{16}\frac{\partial^2}
{\partial r_i
\partial r_j}\,\langle\,\delta u_i \delta u_j (\delta \vec u \cdot
\delta \vec u)^2 \,\rangle\,-\,\frac{1}{4}\frac{D^2}{D\tau^2}
\langle\,(\delta \vec u \cdot \delta \vec u)^2
\,\rangle\,\nonumber \\
&&\,+\,\langle\,2\delta u_i \delta u_j
\partial_i\tilde{p}\partial'_j\tilde{p}'\,+\, \delta u_i \delta u_i
\partial_j\tilde{p}\partial'_j\tilde{p}'\,\,-\frac{D}{D\tau}\langle\,
(\partial_i\tilde{p}+\partial_i'\tilde{p}')\delta u_i \delta \vec
u \cdot \delta \vec u \, \rangle\,
 \nonumber \\& &+\,\frac{1}{2}\partial_{r_\ell}\langle \, \delta u_\ell
\delta \vec u \cdot \delta \vec u \delta u_i(\partial_i
\tilde{p}-\partial'_i\tilde{p}')\,\rangle~~.
\end{eqnarray}

We emphasize that apart from terms negligible in the limit $\nu
\rightarrow 0$ the above equation is exact (the neglected terms
have been evaluated.) To complete our discussion we must consider
the noise and pressure terms in Eq. (\ref{nsbasic}). We first 
note
that
$$\langle
\epsilon_{ii}\epsilon'_{jj}\rangle\,=\,\langle\epsilon\epsilon'\rangle\,+\,\nu
\nabla_r^2\langle \epsilon
\tilde{p}'+\epsilon'\tilde{p}\rangle\,+\,\nu^2\nabla_r^2\nabla_r^2\langle
\tilde{p}\tilde{p}'\rangle$$ which is equal to $\langle \epsilon
\epsilon'\rangle$ in the inertial range. We expect 
the diagonal elements of $\langle \partial_i u_\ell
\partial_j u_\ell \rangle$ to yield the most singular terms 
and these survive in the $\nu \rightarrow 0$ limit which
cuts off the short-distance singularities. With this observation 
we see that only the terms with $i=j$ contribute in
$\langle\epsilon_{ij}\epsilon'_{ij}\rangle$; using the isotropy of
the turbulent state  we find that the left-hand side 
of Eq.(\ref{nsbasic}) reduces to
$(10/3)\,\langle \epsilon \epsilon'\rangle$ in the inertial range.
The noise terms can
be evaluated  and the
non-trivial term in the $\tau \rightarrow 0^+$ limit is precisely
$(10/3)\langle \epsilon \rangle^2$. Thus the left-hand side 
of Eq.
(\ref{nsbasic}) is $(10/3)[\langle\epsilon\epsilon'\rangle\,-\,
\langle\epsilon\rangle^2]$  in the inertial range. Finally, we
observe that the pressure terms depend upon derivatives of the pressure
$\partial_ip$. The kernel in the inversion of
$\nabla^2p\,=\,-\partial_iu_j\partial_ju_i$ is Coulombic and long-ranged; 
however, the pressure-derivative terms can be written with a kernel which is
dipolar and hence, shorter-ranged leading to more convergent 
integrals. Thus writing the 
pressure contributions in
terms of velocities 
one  concludes plausibly  that these contributions will
not be more dominant than those due to the terms
$D^2S_4/D\tau^2$ and $\partial^2S_6/\partial r^2$. 

We discuss our results next. The behavior of $\langle \epsilon
\epsilon'\rangle$ is thus determined by the two terms in the limit
$\tau\,=\,0$,  analogous to those in the Burgers equation. The
term $\partial_{r_i}\partial_{r_j} \langle \delta u_i \delta u_j
(\delta \vec u \cdot \delta \vec u)^2 \rangle$ yields
$\mu_1\,=\,2\,-\,\zeta_6$. The dynamical term, $D^2/D\tau^2
\langle \delta \vec u \cdot \delta \vec u)^2 \rangle$, yields the
identification $\mu_2\,=\,z''_4\,-\,\zeta_4$. 
This provides a transparent derivation directly from the 
Navier-Stokes equation of the two dominant 
exponents characterizing dissipation-rate correlations, one 
which depends purely on the
static structure function exponent ($2-\zeta_6$) and the 
other which involves dynamical behavior ($z''_4-\zeta_4$).

We proceed further, motivated by a similar strategy in the 
theory of phase 
transitions\cite{fisher}, and make an {\em
ansatz} that the two terms are equally dominant. This leads to 
the identification\cite{ansatz}
\begin{equation}2-\zeta_6\,=\,z''_4\,-\,\zeta_4~~.\end{equation}
This relation connects multifractality 
in spatial correlations to multifractality in temporal 
correlations.\cite{lpp} We then obtain the Kolmogorov result 
for the leading dissipation rate correlation exponent, 
$2-\zeta_6$. 

An earlier paper by us\cite{pof} provided a justification for the
two exponents for the dissipation rate correlations 
using  detailed equations  for various 
structure functions explicitly, identifying terms in
different equations and invoking plausible comparisons 
with the Burgers equation result. We
note that other relations such as
$\mu\,=\,2\zeta_2\,-\,\zeta_4$ have been proposed in the
literature.\cite{nelkin}  Other derivations of
 the Kolmogorov relation include one based on fusion rules for equal
 time multipoint correlation functions by L'vov and Procaccia\cite{lp1} who
 also pointed out another scenario which yields
$\mu\,=\,2\zeta_2\,-\,\zeta_4$
 as in ref. \cite{nelkin}. 

We point out that in our equation the subtracted
correlation, $\langle \epsilon \epsilon'\rangle-\langle\epsilon\rangle^2$, 
arises naturally.
In the inertial range we expect the term $(L/r)^\mu$ to dominate over the
constant $\langle \epsilon\rangle^2$ term. However, this might
require $r \,<<\,L$ since $\mu$ is small and this renders the
experimental determination of $\mu$ more difficult. However, for
$r\,<<\,L$ whether one subtracts or not one should obtain the
correct exponent  as expected from these theoretical considerations, and
this is clear in Ref. \cite{sreeni2}.\\

Acknowledgments: One of us (C.J.) thanks M. E. Fisher for a stimulating
discussion. We are grateful to 
K. R. Sreenivasan for helpful comments on the
manuscript. \\


\begin{thebibliography}{99}
\bibitem{frisch43} See page $43$ of Frisch in Ref. $1$. For a  smooth
functional, ${\cal
F}$, of $\hat{u}(k)$ that depends on a Gaussian random noise,
 $\hat{f}(k)$, one has
$$\langle\,\hat{f}(k,t) \hat{{\cal F}}[\hat u(q,t)]\,\rangle\,=\, 
\frac{1}{2} \hat{D}(k)
\langle\,\delta \hat{{\cal F}}/\delta \hat{u}(-k,t) \,\rangle$$
where $D(k)$ is the variance of the noise.
\bibitem{ansatz} This ansatz relating dynamical and static 
exponents is supported by
an exact equation for $S_4$ in the stochastic Burgers problem
which we have derived:
$$\frac{D^2S_4}{D\tau^2}\,=\,\frac{1}{20}\frac{\partial^2 S_6}{\partial r^2}
\,-\,6\frac{D}{D\tau}\langle\,\delta \epsilon  (\delta
u)^2\,\rangle\,+\,\cdots\cdots~~~.$$ This makes the identification 
of $\zeta_6-2$ with $\zeta_4-z_2''$  plausible 
with both exponents being 
determined by the correlation of the dissipation rate with
velocity differences. A similar equation can also be 
derived  for the Navier-Stokes problem.
\end{thebibliography}
\end{document}